\documentstyle[twocolumn,aps,epsf]{revtex}

\begin{document}

\twocolumn[
\hsize\textwidth\columnwidth\hsize\csname@twocolumnfalse\endcsname
\draft

\title{Effect of controlled disorder on  
quasiparticle thermal  transport in Bi$_{2}$Sr$_{2}$CaCu$_{2}$O$_{8}$}
\author{S. Nakamae, K. Behnia and L. Balicas\footnote{On leave from Centro de F\'{\i}sica, 
Instituto Venezolano de Investigaciones Cient\'{\i}cas, Venezuela}}
\address{Laboratoire de Physique Quantique(UPR 5 CNRS), ESPCI, 
75005 Paris, France}
\author{F. Rullier-Albenque}
\address{Service de Physique de l'Etat Condens\'e, CEA-Saclay, 91011 Gif-sur-Yvette, France}
\author{H. Berger}
\address{D\'epartment de Physique, Ecole polytechnique F\'ed\'erale de Lausanne, CH-1015 
Lausanne, Switzerland}
\author{T. Tamegai}
\address{Department of Applied Physics, University of Tokyo, Hongo, Bunkyo-ku, Tokyo, 
113-8656, Japan}
\date{December 4, 2000}
\maketitle
	
\begin{abstract}
Low temperature thermal conductivity, $\kappa$, of  optimally-doped Bi2212 
was studied before and after the introduction of point defects by electron 
irradiation. 
The amplitude of the linear component of $\kappa$ remains unchanged, confirming the
 universal nature of heat transport by zero-energy quasiparticles. The induced decrease
 in the absolute value of $\kappa$ at finite
temperatures allows us to resolve a nonuniversal term in $\kappa$
due to conduction by finite-energy quasiparticles. The magnitude of this term provides
an  estimate of the quasiparticle lifetime at subkelvin temperatures.
\end{abstract}

\pacs{}
] Fifteen years after their discovery, high-T$_{c}$ superconductors continue
to attract significant attention and provoke intense debate\cite{orenstein}.
One central issue is the extent of validity of the Fermi-liquid picture for
describing the electronic excitations in these systems. The properties of
the metallic state (even at optimal doping) appear to remain beyond such a
picture. But, are there well-defined quasiparticles ($qp$) deep in the
superconducting state as suggested by ARPES measurements\cite{kaminsky}? And
if yes, at which energy scale do they break down? To answer these questions,
low-energy excitations in the superconducting state are under intense
scrutiny\cite{durst,chiao}.

Low temperature thermal conductivity, $\kappa $, has proven to be an
instructive probe of such excitations. A non-vanishing linear term in
thermal conductivity of optimally-doped Y123 for $T\rightarrow 0$~ 
was the first solid evidence for a finite density states of
nodal quasiparticles at zero energy \cite{taillefer}. 
Moreover, as expected for the case of a 
$d$-wave gap\cite{lee,graf}, the amplitude of this term was found to be
universal; {\it {i.e.}} independent of impurity concentration\cite{taillefer}%
. Recently, Durst and Lee \cite{durst} showed that, regardless of
Fermi-liquid corrections, this amplitude is intimately related to the fine
structure of the superconducting gap in the vicinity of the nodes.
Subsequently, Chiao {\it {et al.}}\cite{chiao} found a quantitative
agreement between this gap structure as deduced from thermal conductivity and
the one directly observed by ARPES studies\cite{mesot} in optimally-doped
Bi2212. On the other hand, data on electronic specific heat \cite
{moler,wright,wang} and penetration depth \cite{bonn,slee} are limited to T\ 
%TCIMACRO{\TEXTsymbol{>}}%
%BeginExpansion
\mbox{$>$}%
%EndExpansion
2K and probe excitations within a higher energy interval where the density
of states is a linear function of energy. The temperature dependence of
electronic specific heat in Y123 is in agreement with theoretical
expectations for the gap structure near the nodes\cite{chiao,wang}. The same
holds for the thermal variation of superfluid density obtained by
penetration-depth studies assuming a finite Fermi-liquid correction to
charge transport\cite{mesot}. Observing these remarkable quantitative
agreements, Chiao {\it {et al.}}\cite{chiao} argued that this is a strong
indication for the validity of Fermi-liquid treatment of the nodal
quasiparticles. However, one important parameter of this picture, the
impurity bandwidth, $\gamma $, the energy scale below which the density of
states becomes constant, is yet to be measured
experimentally.

Despite such marked accomplishment by the quasiparticle picture to treat the
low-energy nodal excitations, numerous challenges remain. One is
the absence of a linear term in low-temperature thermal conductivity of
underdoped stoichiometric YBa$_{2}$Cu$_{4}$O$_{8}$\cite{hussey}.
As there is no obvious reason to suppose that the structure of the
superconducting gap in this system is radically different from that of the
Y123 parent compound, this result suggests that somewhere in the
underdoped regime, the Fermi-liquid picture might break down. 
Another issue
is the effect of superconductivity on $qp$ lifetime. Transport
studies in both Y123\cite{bonn2,krishana} and Bi2212\cite{slee} have
documented a steep increase in the scattering time of quasiparticles below $%
T_{c}$. According to recent ARPES measurements by Valla {\it et al.} \cite
{valla} on Bi2212, however, the $qp$ lifetime is not affected by
the onset of superconductivity. This surprising discrepancy remains to be
explained.

Here, we present a study of subkelvin thermal conductivity in
optimally-doped Bi$_{2}$Sr$_{2}$CaCu$_{2}$O$_{8}$before and after the
introduction of vacant and interstitial sites by electron irradiation. We
found that the zero-energy $qp$ conductivity remains constant despite the
introduction of pair-breaking defects, while there is a substantial decrease
in $qp$ thermal conductivity, $\kappa_{qp}$, at finite temperatures. 
This allows us to resolve the component of heat transport arising from finite-energy
quasiparticles. 

Two single crystals of Bi$_{2}$Sr$_{2}$CaCu$_{2}$O$_{8}$ were used in this
study with typical dimensions of 1.0 x 0.3 x 0.02 mm$^{3}$. Thermal
conductivity was measured with a conventional two-thermometer-one-heater
set-up. Point defects were introduced by exposing the sample to a 2.5 MeV
electron beam created by Van de Graaf accelerator at the Laboratoire des
Solides Irradi\'{e}s in Palaiseau, France. Fig. 1 show the effect of
electron irradiation on electrical resistivity and the corresponding low
temperature thermal conductivity of the two samples. As seen in the two
insets, irradiation leads to a decrease in the critical temperature and an
increase in normal-state resistivity, $\rho$, as reported in previous
electron-irradiation studies\cite{flo}. Both pristine samples have a
transition temperature ($T_{c}$ =92K) and $d\rho/dT$ 
($\rho=\rho_{o}+ bT \mu \Omega$~cm, with $b=1.5\mu \Omega$~cmK$^{-1}$)
characteristic of optimally doped Bi2212 crystals\cite{watanabe}. 
In the case of sample $a$ ($\rho _{o}\sim $60 $\mu \Omega)$, an irradiation flux 
of 6.0$\times$10$^{19}$ e$^{-}$/cm$^{2}$ leads to a fourfold increase in 
residual resistivity ($\rho_{0}\sim $240 $\mu\Omega$ cm) 
and a 20 K decrease in $T_{c}$ (73K).
Moreover, a curvature appears in the temperature-dependence of resistivity
which may suggest a doping change. 
In order to explore such possibility, 
we have measured the room-temperature thermopower, $S(300K)$, 
which is a well known indicator of doping levels in high-$T_c$ superconductors\cite{obertelli}. 
The determined values of post-irradiation samples $a$ and $b$ (after the third irradiation) 
are -$0.1\pm $ 0.1 and $5.0\pm$ 0.1 $\mu $ V/K.  
The values of
corresponding un-irradiated crystals are -0.8 $\pm$ 0.4 and 2.4 $\pm$ 0.1 $\mu $ V/K, 
thus in both samples irradiation appear to have caused a slight ``under-doping". 
According to the study on $S(300K)$ of Bi2212
\cite{obertelli}, $\Delta T_c$ due to a possible variation in the doping level 
here are 1.8 K and 4.5 K for samples $a$ and $b$, respectively.  
These changes are considerably
smaller than $\Delta T_c$ due to the pair-breaking effect from electron irradiation,
20 and 50 K, and thus we will assume that the effect on $qp$ transport arising from a small
change in the doping level is negligible in comparison.
It must be noted here that the thermal conductivity of sample $b$ was not measured before
irradiation. Thus in the following discussions, we will concentrate on sample $a$
and will use the data on the sample $b$ only as supplementary evidence. 

The main panels of Fig. 1 present the low temperature thermal conductivity data.
The persistence of a sizeable phonon contribution, $\kappa _{ph}$, to the total heat
transport complicates the determination of $qp$ conductivity. As $%
T\rightarrow 0$, $\kappa _{ph}$, is expected to
enter the boundary scattering (or ballistic) regime, 
where phonon mean-free-path is limited
by the dimensions of the crystal and $\kappa _{ph}$ becomes proportional to $%
T^{3}$. Above this regime, $\kappa _{ph}$ should increase more slowly due to
the decrease in phonon mean-free-path. Therefore, by plotting $\kappa /T$%
~against $T^{2}$, one can resolve the zero-intercept, $\kappa _{00}/T$,
corresponding to the residual linear component of $\kappa$ that
is a signature of residual $qp$ conductivity. As seen in fig. 1b, there is a
finite $\kappa _{00}/T$ term in the thermal conductivity of sample $a$ both
before and after irradiation. The magnitude of this term shows little or no
change in spite of the drastic increase in the number of defects. 
Note that the original contacts were kept throughout the successive measurements
for each crystal, the sizeable uncertainty in $\kappa$
geometric factor does not hamper the effect of disorder. 
If we assume a linear extrapolation of the normal-state resistivity, 
the scattering rate in sample $a$ becomes four times larger after irradiation.
If we further take the enhancement in the $qp$ lifetime
below T$_{c}$ into account, the induced increase in impurity scattering rate 
becomes even larger. 
The amplitude of $\kappa _{00}/T$ found in sample $a$ (0.16 $\pm $ 0.03 mW/K$^{2}$cm)
is similar to that of sample $b$ (0.13 mW/K$^{2}$cm) 
as well as to those reported in previous studies
(0.14-0.15 mW/K$^{2}$cm) \cite{stockholm,chiao}. 
Theory predicts an increase in $\kappa _{00}/T$
at sufficiently high levels of impurity concentration due to
the impurity-induced change in T$_{c}$ \cite{maki}. 
Therefore, considering that the sample $b$ after the third irradiation is nearly twenty times 
dirtier than the pristine sample $a$ (judging from their respective $\rho_o$ values), 
the ``universal'' character found in the values of $\kappa_{00}/T$ 
among all samples here is quite surprising. 
A quantitative comparison to the fine structure of the superconducting energy 
gap around the nodes can then be
performed by inserting the value of $\kappa_{00}/T$ into the following equation 
\cite{durst}: 
\begin{equation}
\kappa_{00}/T=\frac{{k_{B}}^{2}}{3\hbar }\frac{{v_{F}}}{{v_{2}}}\frac{n}{d}.
\label{1}
\end{equation}
where $v_{F}$~and $v_{2}$~ are $qp$ Fermi velocity and gap
velocity at the nodes in the gap. 
One then obtains $\frac{{v_{F}}}{{v_{2}}}=21\pm 3$~which is close to
the ratio obtained by ARPES measurements\cite{mesot}.  
The same conclusion has been
previously drawn by Chiao {\it et al.}\cite{chiao}.

Before discussing the finite temperature $qp$ conductivity further, let us
focus on phonon conductivity. 
For the sample $a$, if one can assume that the system
enters the ballistic regime for $T < 0.25$~K, the extracted cubic term
for phonon conductivity is $\kappa _{ph}/T^{3}\sim 2.6$mW/K$^{4}cm$. 
This can be compared to the theoretically expected value using $\kappa _{ph}=1/3\beta
\mbox{$<$}v_{ph}
\mbox{$>$}
%EndExpansion
l_{ph}T^{3}$, where $\beta $ is the phonon specific heat coefficient
and $v_{ph}$ is the average sound velocity. Taking the transverse dimensions
of our crystal along with the available values for $\beta
=0.0095$mJ/K$^{4}$cm$^{-3}$\cite{junod} and $v_{ph}=3600$ms$^{-1}$\cite{boek}, 
$\kappa _{ph}/T^{3}$ should be 5.9 $mW/K^{4}cm$, approximately
twice the value determined from our data.
One possible source for this discrepancy is that our measurement stops above
the onset temperature of phonon ballistic conductivity. In the study reported
by Chiao {\it et al.}, 
%covering temperatures as low as 80 mK, 
the ballistic
regime appears only below 130 mK. In this case, a downward curvature of the 
$\kappa /T$ curve would lead to a slight decrease in the finite intercept and
the estimated $\kappa_{00}$ would become 0.14 mW/K$^{2}$cm (see Fig. 2a)
but remains within our experimental uncertainty.

Another important aspect of Fig. 1b is a sizable
irradiation-induced decrease in $\kappa (T)$  for the entire temperature
range (0.13 K 
\mbox{$<$}%
T 
\mbox{$<$}~0.9K). 
We begin by noting that the typical wavelength
of acoustic phonons may be estimated to vary as $\lambda
_{ph}=hv_{ph}/k_{B}T=173$nm/K. Thus, at subkelvin temperatures, the spatial
extension of lattice vibrations is more than two orders of magnitude larger
than the size of introduced point defects and thus $\kappa_{ph}$ should 
not be affected by irradiation.
For this reason, the observed decrease must 
be exclusively due to an increase in the $qp$-defect scattering rate.
However, the $qp$ conductivity is expected to become {\it {independent}} 
of impurity scattering rate for $T%
\mbox{$<$}%
\gamma$, the
impurity bandwidth.  Such condition is finally met in sample $b$ after
the second and the third irradiation where $\kappa$ is virtually
unchanged (see the lower panel of Fig.1). 
In relatively cleaner samples, the marked decrease in $\kappa$~infers that
$\gamma$ must lie below our range of measurements. 

Fig. 2 depicts the change in $\kappa (T)/T$ of sample $a$ before and after
irradiation. $\Delta \kappa /T$ is
linear in $T$, revealing a quadratic temperature dependence. 
The most plausible
explanation for this result is to concede that $qp$ conductivity of
the pristine sample includes a quadratic term which is heavily diminished by
irradiation. A fit to the data of Fig. 2 for the whole temperature range
yields $\Delta \kappa =aT+bT^{2}$ with $a=$0.005mW/K$^{2}$cm and $
b=$0.19mW/K$^{3}$cm. The small size of the linear intercept indicates again
that irradiation has left the linear term of $qp$ conductivity
intact. We now compare the amplitude of the term $b$ with what is expected from
the finite-energy $qp$ contribution. For energies
exceeding $\gamma$, the density of states 
%in $d$-wave superconductors 
varies linearly with energy\cite{durst,chiao}:

\begin{equation}
N(E)=\frac{2}{\pi \hbar ^{2}}\frac{1}{v_{F}v_{2}}E.  \label{3}
\end{equation}
This leads to a quadratic temperature dependence of specific heat\cite{chiao}%:

\begin{equation}
C_{e}=\frac{18\zeta (3)}{\pi }\frac{k_{B}^{3}}{\hbar ^{2}}\frac{n}{d}\frac{1%
}{v_{F}v_{2}}T^{2}  \label{4}
\end{equation}
where $\zeta (3)\simeq 1.20$ is a numerical factor.~Thermal conductivity of
these excitations can be estimated via kinetic theory; $\kappa _{e}$=$\frac{1%
}{3}C_{e}v_{F}^{2}$ $\tau_{e}$. Here, $\tau _{e}$ is the electronic scattering time. 
The temperature dependence of $\tau_{e}$
governs the temperature dependence of electronic thermal conductivity.
The relative weight of this term to the universal linear term can be
estimated to be:

\begin{equation}
\frac{\kappa_{e}}{\kappa_{00}}=\frac{18\zeta (3)k_{B}\tau_{e}T}{\pi \hbar}  
\label{(5)}
\end{equation}
Note that while $\kappa_{00}/T$ is universal, $\kappa_{e}$ is not: its
magnitude decreases with increasing disorder. The size of the
quadratic term coefficient appearing in $\Delta \kappa (T)$~($b=$0.19mW/K$^{3}$cm) 
corresponds to an electronic scattering time of 1.3 ps. On the
other hand, $qp$ lifetime can be estimated from 
$\rho =\frac{m^{\ast}}{ne^{2}\tau_{e}}$ and $\omega_{p}=(4\pi ne^{2}/m^{\ast})^{1/2}$,
where $\omega _{p}$ is the Drude plasma frequency. 
Using $\omega_{p}=$~1.1eV \cite{romero}, 
$\rho (100K)=200$$\mu \Omega cm$ and $\rho_{0}=60\mu\Omega cm$, 
one finds $\tau_{e}(100K)\simeq$~0.05 ps and $\tau_{e}(0K)\simeq$~0.2 ps, 
provided that the
scattering rate is also linear in temperature for $T< T_{c}$. 
Our results suggest a significant increase in $qp$ lifetime, 
and thus, a substantial reduction in scattering rate below $T_{c}$, 
in agreement with microwave conductivity data\cite{slee}. 
It is instructive to compare the magnitude of $\tau_{e}$ in Bi2212
and Y123. In the latter system, $\tau_{e}$ of high-quality crystals is
of the order of 7 ps \cite{bonn2} and is reported to approach 20 ps \cite{hosseini} 
in BaZrO$_{3}$-grown crystals. 
For Bi2212, recent study on the complex conductivity
by Corson {\it et al.} showed the $qp$ lifetime to approach $\sim$1 ps
\cite{Corson} at low temperatures, 
similar to the value calculated independently from our measurements.
%We note that our estimation of the
%quasiparticle lifetime is in agreement with the low-temperature value ($\sim 
%$1 ps) reported in a recent study of complex conductivity in Bi2212 thin
%films \cite{Corson}. 
Further, assuming the scattering rate to remain constant\cite{hosseini}, 
one can estimate the size of $\kappa _{e}$ upward in temperature.
The deduced value at 5K ($\sim $4.8 mW/Kcm) can be compared via the
Wiedemann-Franz law to the available data on charge conductivity which is
limited to T %
%TCIMACRO{\TEXTsymbol{>}}%
%BeginExpansion
\mbox{$>$}%
%EndExpansion
5K \cite{slee}. 
This yields $\frac{\kappa _{e}}{\sigma _{1}T%
}(5K)\sim 21.3$ nW/$\Omega$ K, very close to the Sommerfeld value (L$_{0}$%
=24.5 nW/$\Omega$ K). 
Thus, our interpretation appears to be consistent with
what is known from charge conductivity.

In spite of this apparent consistency, this analysis cannot accommodate
the broader theoretical picture of $qp$ transport in d-wave superconductors
elaborated during the recent years. First, a linear N(E) implies an
energy-dependent $\tau _{e}$.  In the unitary limit, for example, this
leads to a cubic (instead of a quadratic) behavior for finite-energy
$qp$ transport for $T >\gamma $,\cite{graf}. Second, the size
of $\gamma $ strongly depends on the impurity density ($n_{imp}$) as well as
the scattering phase shift $\delta $.  In the unitary ($\delta =\pi /2$)
limit, $\gamma $~increases substantially with increasing scattering rate, $\Gamma$;
$\gamma \sim (\Gamma \Delta _{0})^{1/2}$. 
In the Born ($\delta =$~0) limit, its enhancement with $\Gamma$~is largely
attenuated for $\Gamma \ll T_{c}$; $\gamma \sim \Delta _{0}exp(-\frac{\Delta
_{0}}{\Gamma })$.  Here $\Gamma =\frac{1}{2\tau _{e}}$ and $\Delta _{0}$
is the magnitude of the superconducting gap \cite{lee}. 
Now, with $\tau _{e}\sim $~1 $ps$ (which would
yield $\Gamma \sim 3$K)~ and $\Delta _{0}\sim$ 40meV \cite{renner}, $\gamma $
~may be estimated to be 35K in the unitary limit and virtually zero in the
Born limit. 
Thus, the size of $\gamma$ found in this study is
in sharp contrast with what is expected in the unitary limit. 
A slight deviation from the unitary limit can
produce a linear $\kappa (T)/T$ above its universal value in a limited
temperature range \cite{hirschfeld}, however, preliminary investigations reveals that 
a reasonable phase shift only cannot account for the magnitude of the excess 
conductivity observed here \cite{vekhter}. Clearly, our findings
constitute a challenge to existing theory.

In summary, we have studied the effect of electron irradiation on the low
temperature thermal conductivity of optimally doped Bi$_{2}$Sr$_{2}$CaCu$%
_{2} $O$_{8}$. The quasiparticle contribution to heat transport was found to
contain two distinct components, a linear term associated with zero-energy
quasiparticles and a quadratic term originating from finite-energy
quasiparticles. The linear term remains insensitive to the number of
defects. The magnitude of the quadratic term allowed us to make a new
independent estimation of quasiparticle scattering time at low
temperatures, implying a significant increase in the quasiparticle
transport lifetime below $T_{c}$ as indicated by other probes.
\\

This work was partially supported by a NSF grant (No. INT-9901436) and a
CNRS/CONICIT joint grant (No. 7197). We acknowledge enlightening discussions
with P. J.\ Hirschfeld, N.\ E.\ Hussey, J.\ Lesueur L.\ Taillefer and I.
Vekhter.

\begin{figure}[tb]
\caption{Upper panel: Low-temperature thermal conductivity, $\protect\kappa/T$,
of sample $a$ is plotted as a function of $T^{2}$. The thin lines are
guides to the eye. The thick line represents the expected asymptotic lattice
conductivity at the ballistic regime when the phonon mean-free-path attains
the average sample size (see text).The inset shows the resistivity data on
the same sample.\ Lower panel: Same for sample $b$. Note that the theral
conductivity of this sample in the pristine state was not measured.}
\end{figure}

\begin{figure}[tb]
\caption{Upper panel: The change in the thermal conductivity of the sample
before and after electron irradiation divided by temperature and plotted as
a function of temperature. The straight line is a fit to the data revealing
a negligible intercept and a quadratic variation of $\Delta \protect\kappa
(T)$. Lower panel: Thermal conductivity of the sample before and after
irradiation together with a sketch of electronic and lattice components of
thermal conductivity in the pristine sample.}
\end{figure}

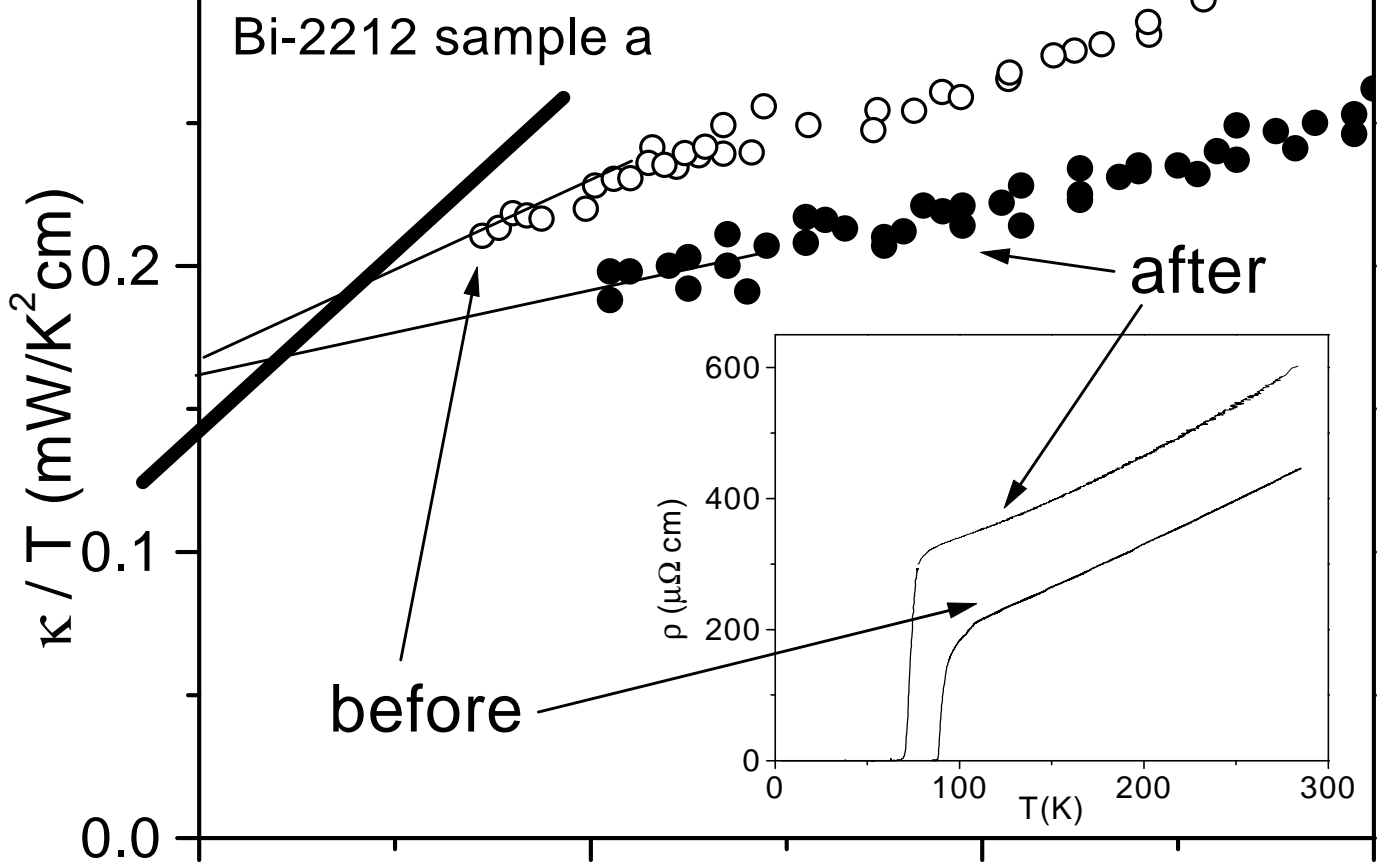
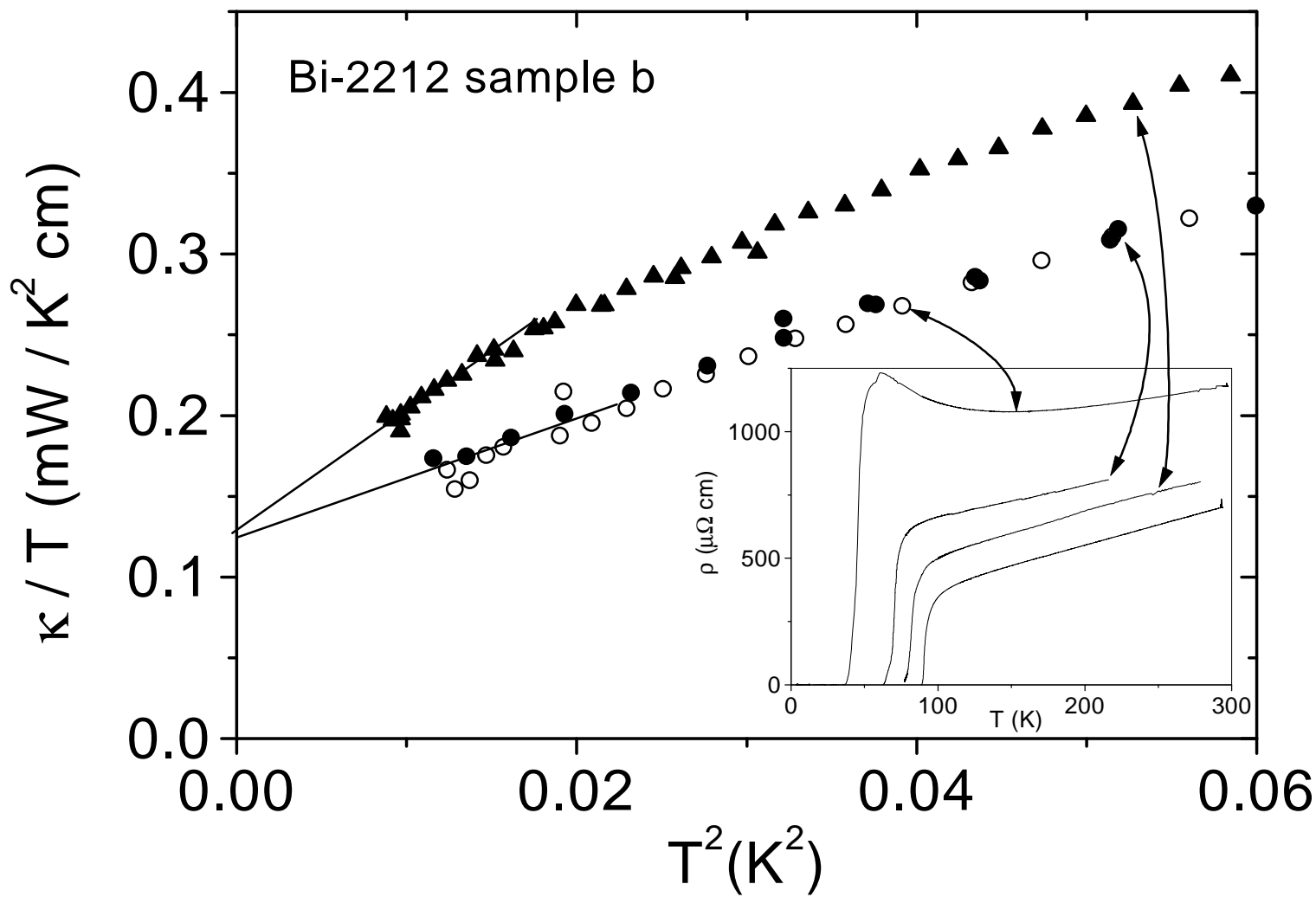
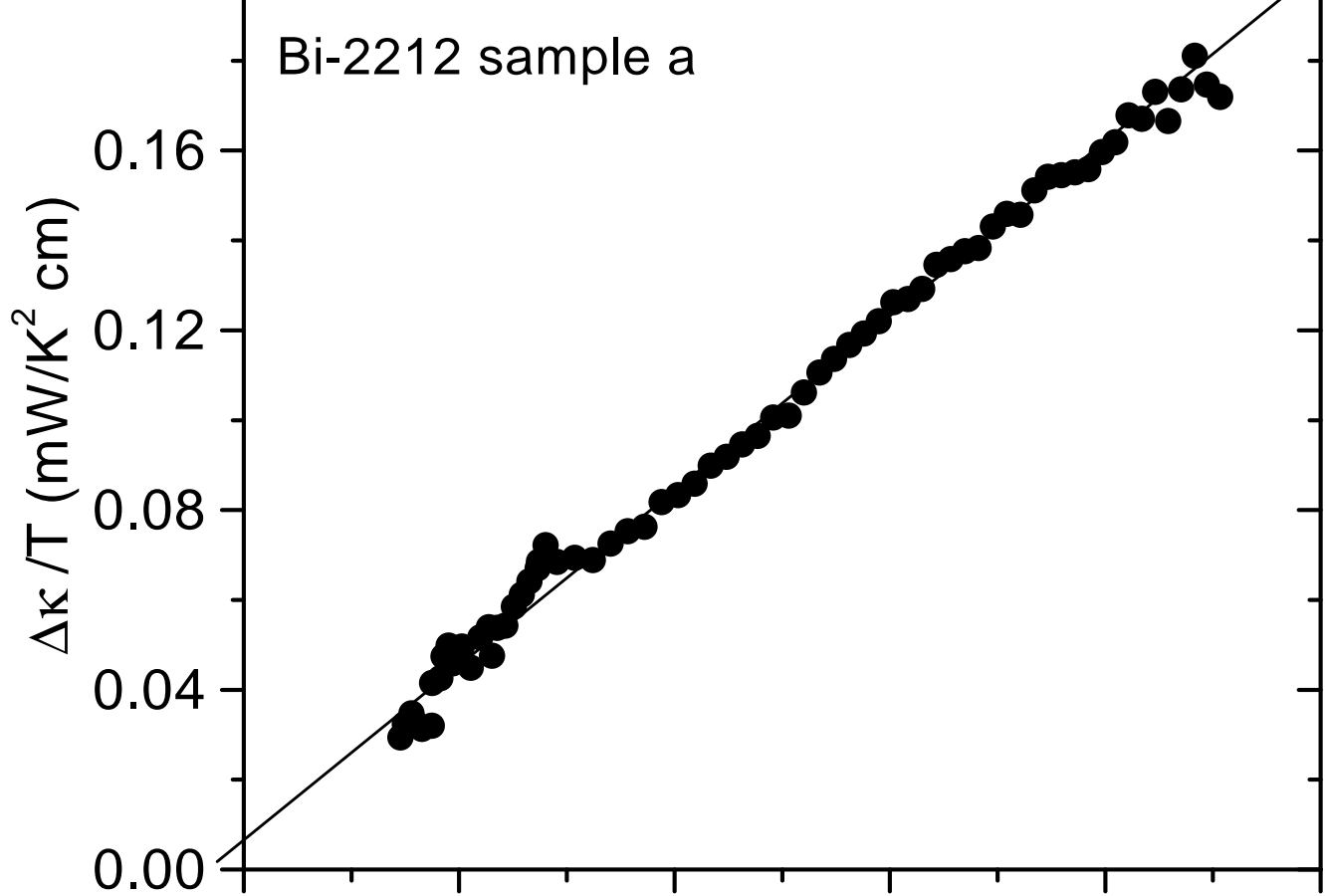
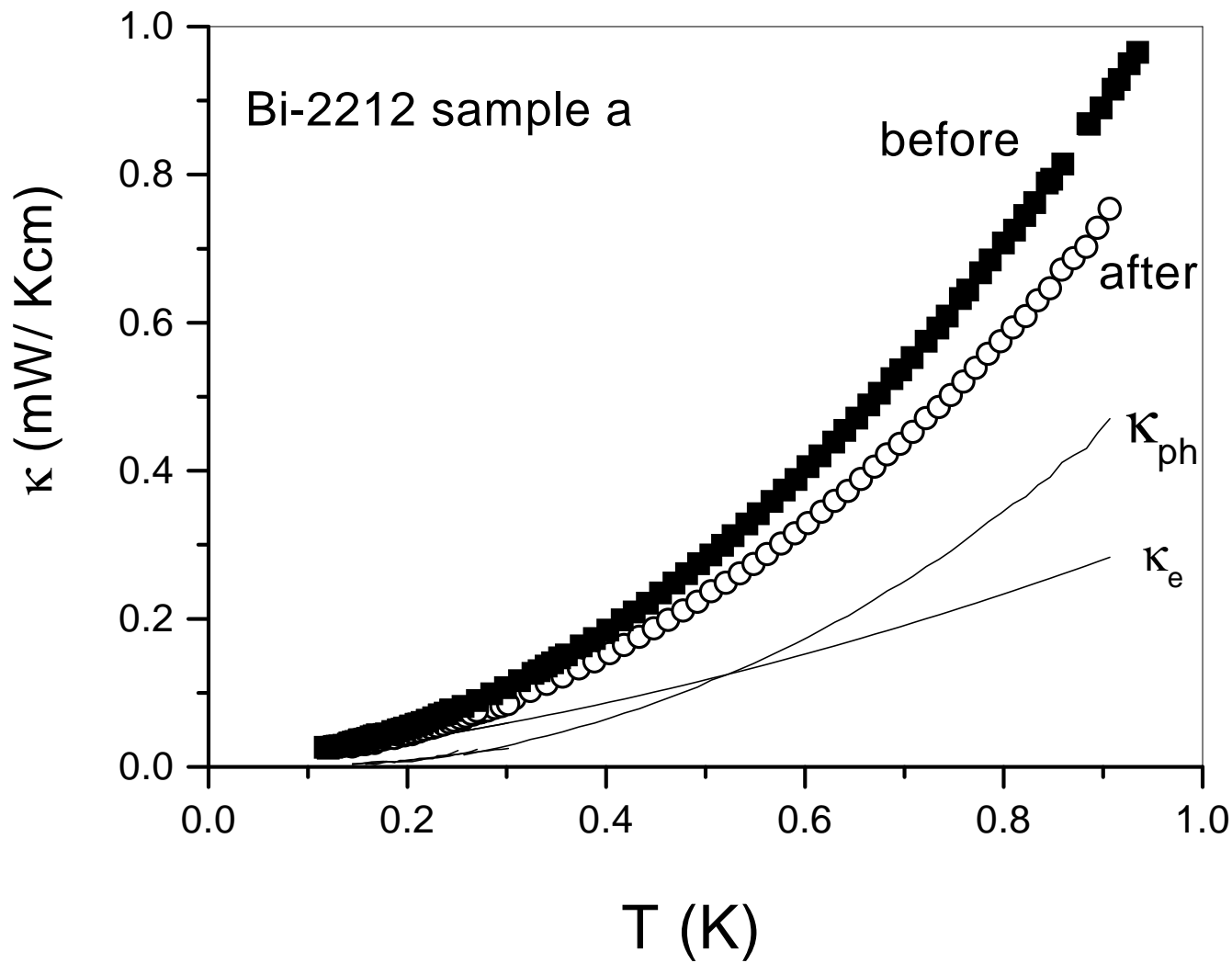
\end{document}